\documentclass{PoS}
\usepackage{lineno}
\title{Search  for tau neutrinos at PeV energies and beyond with the MAGIC telescopes}

\ShortTitle{Search  for tau neutrinos with the MAGIC telescope}

\abstract{The MAGIC telescopes, located at the Roque de los Muchachos
Observatory (2200 a.s.l.) in the Canary Island of La Palma,
are placed on the top of a mountain, from where a window of visibility of about 5 deg in zenith and 80 deg in azimuth is open in the direction of the surrounding ocean. This permits to search for a signature of particle showers induced by earth-skimming cosmic tau neutrinos in the PeV to EeV energy range arising from the ocean. We have studied the response of MAGIC to such events, employing Monte Carlo simulations of upward-going tau neutrino showers. The analysis of the shower images shows that air showers induced by tau neutrinos can be discriminated from the hadronic background coming from a similar direction. We have calculated the point source acceptance and the expected event rates,  for a sample of generic neutrino fluxes from photo-hadronic interactions in AGNs. The analysis of about 30 hours of data taken toward the sea leads to a point source sensitivity for tau neutrinos at the  level of  the down-going point source analysis of the Pierre Auger Observatory, if the AUGER observation time is dedicated to a similar amount by MAGIC.}

\author{\speaker{D. G\'ora$^{1}$}, M. Manganaro$^{2,3}$, E. Bernardini$^{4,5}$, M. Doro$^{6}$, M. Will$^{2,3}$, S. Lombardi$^{7}$, J. Rico$^{8}$, D. Sobczynska$^{9}$, for the MAGIC Collaboration \thanks{https://magic.mpp.mpg.de/acknowledgements\_19\_05\_2017.html}\\
        E-mail: \email{Dariusz.Gora@ifj.edu.pl}}

\author{\\                                                                                                                                                                               
        $^1$Institute of Nuclear Physics Polish Academy of Sciences, PL-31342 Krakow, Poland\\
        $^2$Inst. de Astrofisica de Canarias, E-38200 La Laguna, Tenerife, Spain\\  
        $^3$Universidad de La Laguna, Dpto. Astrof'isica, E-38206 La Laguna, Tenerife, Spain\\                                                                                                                                                               $^4$Deutsches Elektronen-Synchrotron (DESY), D-15738 Zeuthen, Germany\\
        $^5$Humboldt University of Berlin, Institut f\''ur Physik Newtonstr. 15, 12489 Berlin Germany\\
        $^6$Universita di Padova and INFN, I-35131 Padova, Italy\\
        $^7$INAF National Institute for Astrophysics, I-00136 Rome,Italy\\
        $^8$Institut de Fisica d'Altes Energies (IFAE), The Barcelona Institute of Science and Technology, Campus UAB, 08193 Bellaterra (Barcelona), Spain\\
        $^9$University of  L\'odz, PL-90236 Lodz, Poland\\
 }

\FullConference{EPS-HEP 2017, European Physical Society conference on High Energy Physics\\
		5-12 July 2017\\
		Venice, Italy}

\begin{document}

\section{Introduction}
The discovery of an astrophysical flux of high-energy neutrinos by IceCube~\cite{HESE2} was a major step 
forward in the ongoing search for the origin of cosmic rays, since  neutrino emission 
 by usually follows from hadronic interaction in astrophysical accelerators. The  composition of neutrino flux at Earth is
consistent with equal fractions of all neutrino flavors, though with large uncertainty~\cite{icecuflavour,icecuflavour1}. Of particular interest is the identification of $\nu_{\tau}$, which is only expected to be produced in negligible amounts in astrophysical  accelerators, but should appear in the flux detected by IceCube due to neutrino flavor oscillation. Up to now, there has been no clear identification of  $\nu_{\tau}$ at high energies, due to their resemblance with signals induced by $\nu_e$ in ice/water detectors.
However, the detection  of $\nu_{\tau}$  is very important from both the astrophysical and  particle physics point of view. 

A conventional approach for the detection of neutrinos with energies in the PeV range is based on detectors which use large volumes of ice (IceCube) or water (ANTARES).  They sample Cherenkov light from  muons produced by muon neutrinos, or from electron and tau lepton induced  showers initiated by the charged current interactions of electron and tau neutrinos. An alternative technique is based on the proposed observation  of upward going extensive air showers  produced by the leptons originating from neutrino interactions below the surface of the Earth,  the  so-called earth-skimming method~\cite{fargion,bertou}. 

In this paper we study the possibility to use the  MAGIC (Major Atmospheric Gamma Imaging Cherenkov) telescopes to search  for air showers induced by tau neutrinos  ($\tau$-induced showers) in the PeV-EeV energy range. MAGIC is a system of  two  IACTs located at the Roque de los Muchachos Observatory (28.8$^{\circ}$ N, 17.9$^{\circ}$ W), in the Canary Island of La Palma (Spain). They are placed 85 m apart, each with a primary mirror of 17 m diameter. The MAGIC telescopes, with a field of view (FOV) of 3.5$^{\circ}$, have been built  to detect cosmic $\gamma$-rays in the energy range 50 GeV - 50 TeV~\cite{magicperformance}. 

In order to  use MAGIC for tau neutrino searches, the telescopes need to be pointed in the direction of the tau neutrinos escaping first from the Earth crust and then from the ocean, i.e. at the horizon or a few degrees below. In such cases the telescopes can monitor a large volume in  their FOV resulting  in a space angle area (defined as the intersection of the telescopes  FOV  and  the sea surface) of about a few km$^2$.  In~\cite{upgoing_magic}, the effective area for up-going tau neutrino observations with the MAGIC telescopes was calculated analytically and found to reach  $\sim 10^3$ m$^2$ (at 100 TeV) and  10$^5$ m$^2$ (at 1 EeV) for an observation angle of about 
1.5$^{\circ}$ below the horizon, rapidly diminishing with higher inclination.

From an observational point of view, it is worth noting that the time that can be dedicated to this kind of observations  almost does not interfere  with regular MAGIC gamma-ray observations of the sky, because the sea can be pointed even 
 in the presence of optically thick clouds above the MAGIC site (high clouds). 
In fact, high-altitude clouds prevent the observation of gamma-ray sources
but still allow pointing the telescopes to the horizon. For the MAGIC site there
are up to about 100 hours per year available,  when high clouds are present. 

\section{MAGIC observations and Monte Carlo simulations}

\vspace{-0.25cm}
The MAGIC telescopes have taken  data  at very large zenith angles ($85^{\circ} <\theta < 95^{\circ}$) in  the direction of the sea (seaON), slightly above the sea (seaOFF) and towards  the Roque de los Muchachos  mountain. The seaON data were taken at a zenith angle of $\theta=92.5^{\circ}$ while seaOFF  was taken at  $\theta=87.5^{\circ}$.  The  rate  of stereo  seaOFF events is about 27 times  larger   ($\sim$ 4.6 Hz) than  for seaON ($\sim$0.17 Hz) observations. Thus   9.2 hours of seaOFF data  provide  high-statistics background estimates for  about  30 hours of seaON data. It is  worth  to mention, that  the  91\% of the data were taken during nights characterized by optically thick high cumulus clouds, when normal $\gamma-$ray observations are usually worthless. 
\begin{figure*}[t!]{}
\includegraphics [width=0.49\textwidth,height=6cm]{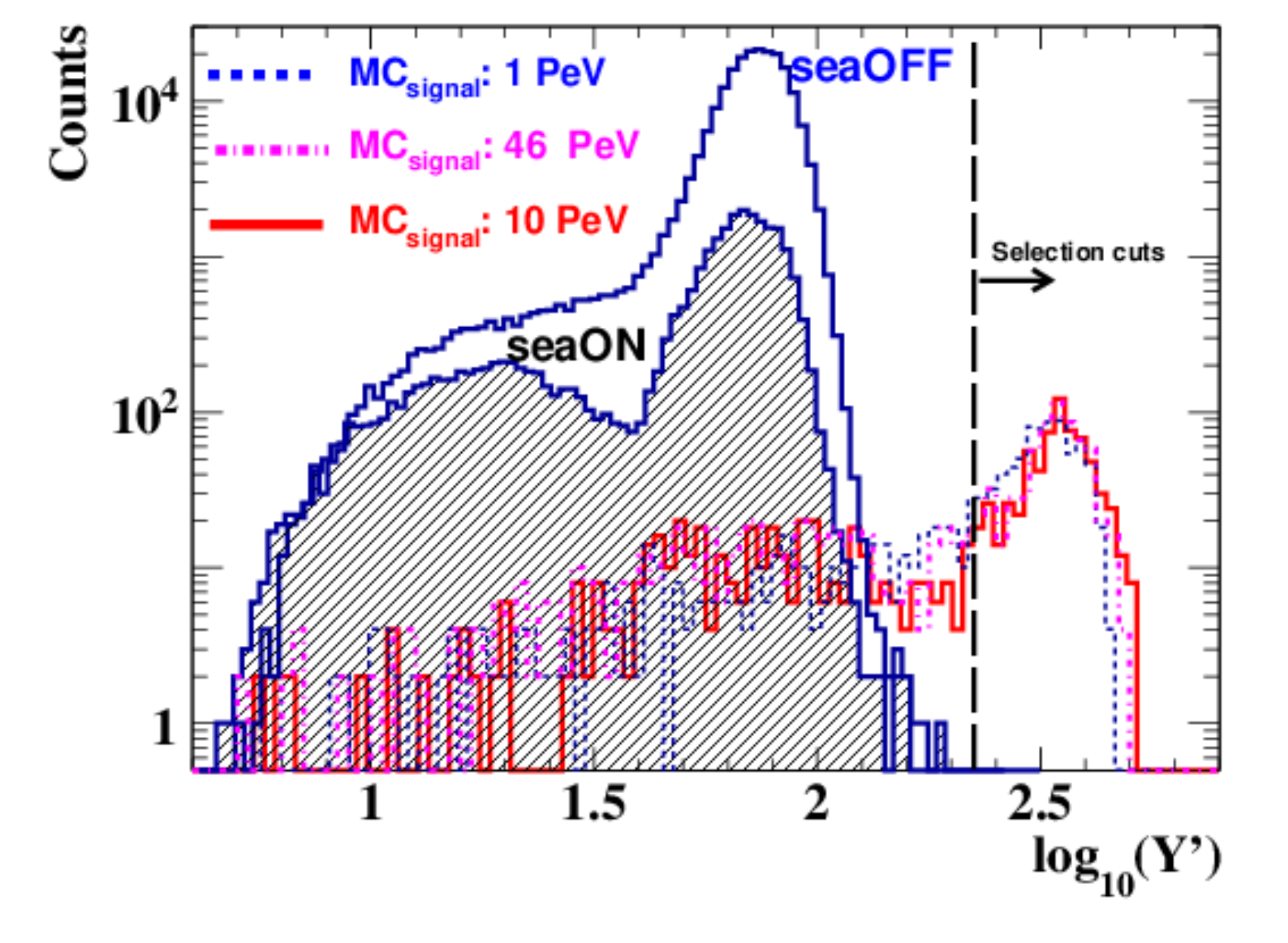}
\includegraphics [width=0.55\textwidth,height=6cm]{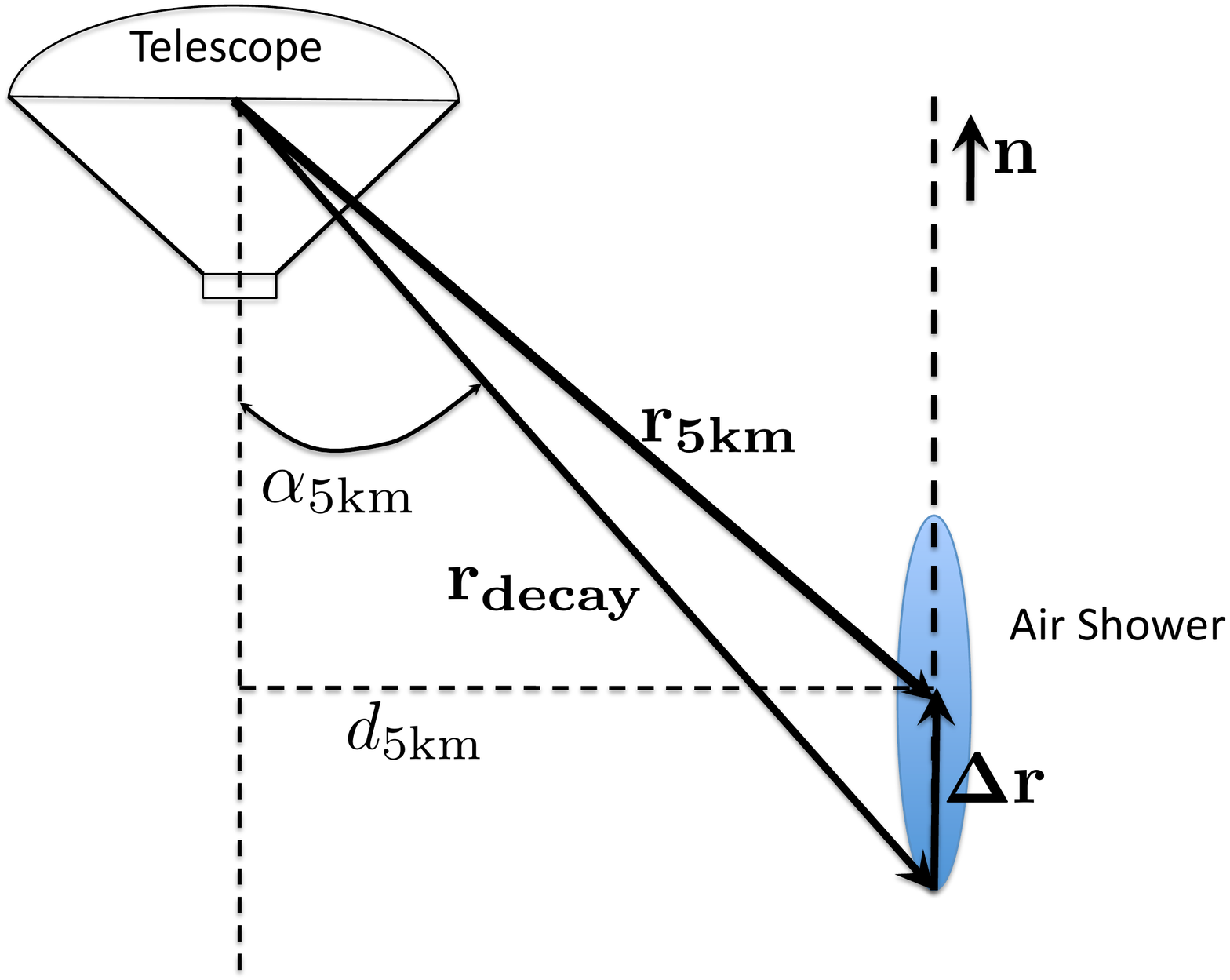}
\caption{ \small Left panel: The one-dimensional distribution of seaON, seaOFF and signal MC. The  coordinate of events were obtained from  the following formula: $\log_{10}(Y{'})=\log_{10}(Size[p.e.])*\cos(\alpha)-\log_{10}(Length[deg])*\sin(\alpha)$, where $\alpha=63.435^{\circ}$.  Note that  above our  selection cut ($\log_{10}(Y^{'})>2.35$)  zero  neutrino candidates are found. For showers with the  larger impact distances (0.3-1.3 km) a slightly  relaxed cut was used:  $\log_{10}(Y^{'})>2.10$; Right panel:  Sketch  illustrates  the FOV cut used in our analysis, see text for more details. } \label{fig::dist}.
\vspace{-0.5cm}
\end{figure*}

In order to study the signatures expected from neutrino-induced showers by MAGIC, a
full Monte Carlo (MC)  simulation chain was set up, which consists of three steps. First, the interaction of a given neutrino flux with the Earth and propagation of the resulting charged lepton through the Earth and the atmosphere is simulated using an extended version~\cite{goraanis} of the ANIS code~\cite{anis}. Second, the shower development of $\tau$-induced 
showers and their Cherenkov light production is simulated with CORSIKA~\cite{corsika}. The results of the CORSIKA simulation are used as inputs for the last step, i.e. the simulation of atmospheric extinction and the MAGIC detector response~\cite{mars}, see \cite{tauicrc2017} for more details.

Each simulated event recorded and calibrated consists of a number of photoelectrons per camera pixel, which has been extracted using a sliding window algorithm~\cite{magicperformance}.
 In order to get rid of pixels whose contents are likely due only to night sky background (NSB)   an image cleaning is performed. The resulting cleaned shower image contains only the pixels  considered to obtain physical information about the shower. The cleaned camera image  is characterized by a set of image parameters introduced by M. Hillas in~\cite{hillas}.  These parameters provide a geometrical description of the images of showers and are used to infer the energy of the primary particle, its arrival direction, and to distinguish between $\gamma-$ray and hadron induced showers. In~\cite{tauicrc2017} we  study  these parameters  for the case of  deep $\tau$-induced simulated showers and compare the corresponding distributions with  data. It is interesting   that only with   two Hillas  parameters i.e.  like $Size$ and  $Length$, we can easily  identify  a region with no background events, and construct a selection cut to identify tau-induced showers, see again~\cite{tauicrc2017} for more details.  Here, we only show   the one-dimensional distribution of seaOFF, seaON and MC signal simulation projected onto the  line perpendicular to our selection cut, see Figure~\ref{fig::dist} (left  panel). As can be   seen  we did not find  any neutrino candidate, if the selection cut is applied  to  all  seaON data. This plots shows that MAGIC can discriminate deep $\tau$-induced  showers from the background of  hadronic showers at high zenith angles.

\section{Monte Carlo estimate of MAGIC acceptance}
As previously mentioned the propagation of a given neutrino flux through the Earth and the atmosphere is simulated using  an extended version of the ANIS code.  Based on these simulations, the detector  aperture/acceptance 
for an initial neutrino energy $E_{\nu_\tau}$ can be  calculated from:
%\begin{eqnarray}
$A^{\mathrm{PS}}(E_{\nu_\tau}, \theta,\phi)  =N_{\mathrm{gen}}^{-1} \times \sum_{i=1}^{N_{\mathrm{FOVcut}}} P_{i}(E_{\nu_\tau},E_{\tau},\theta) \nonumber  
    \times  A_i(\theta) \times T_{\mathrm{eff},i}(E_{\tau},r_{5 \mbox{km}},d,\theta)$,
%\label{aperture}
%\end{eqnarray}
where $\theta$, $\phi$ are the simulated zenith and azimuth pointing angles of the MAGIC telescope,  $N_{\mathrm{gen}}$ is number of neutrino events from the direction $\theta$ and $\phi$. $N_{\mathrm{FOV cut}}$ is the number of $\tau$ leptons with energies $E_{\tau}$ larger than the threshold energy $E_{\mathrm{th}}=1$\, PeV and  with an estimated  position of the shower maximum in the FOV of the MAGIC  telescope, see   Figure~\ref{fig::dist} (right panel) for an illustration of the geometry.

In addition the impact distance $d$ of  the $\tau$-lepton induced showers  is required to be smaller  than 1.3 km. $P(E_{\nu_\tau},E_{\tau},\theta)$ is the probability that a neutrino with energy $E_{\nu_\tau}$ and zenith angle $\theta$  produces a lepton with energy $E_{\tau}$. $A_i(\theta)$ is the physical cross-section area of the interaction volume seen by the neutrino, simulated by a cylinder with radius of 50 km and height 10 km. $T_{\mathrm{eff},i}(E_{\tau},r_{5 \mbox{km}},d,\theta)$ is the trigger and reconstruction/cut efficiency for $\tau$-lepton induced showers with  its estimated position of the shower maximum  at  distance $r_{5 \mathrm{km}}$ from the telescope  and the shower impact distance  $d$. The trigger efficiency depends on the response of a given detector and is usually estimated based on MC simulations,
see \cite{tauicrc2017} for more details.

For  shower energies relevant in this analysis i.e. 1-1000 PeV, the largest longitudinal extension of the shower is  reached
approximately  after  600 g/cm$^2$~\cite{xmax}~\footnote{In this approach shower-to-shower fluctuations are not included, which 
for  a single shower can lead to the  slightly different $\Delta r$ . }. Assuming that  tau-induced showers develop in the densest part of the atmosphere with the near ground density of  air of about  $\rho_{air}=0.0012$ g/cm$^3$, this  depth  interval corresponds to about $\Delta r=5$ km. Thus, a position of the shower maximum can be estimated from the simple formula:   $ \vec{r}_{ 5 \mathrm{km}}=\vec{r}_{\mathrm{decay}}+\Delta r \cdot  \vec{n}$, where   $\vec{r}_{decay}$  is the vector pointing  from the telescope  to the  decay vertex of the tau lepton and $\vec{n}$ is  the unit vector 
describing the direction of the shower. In our FOV cut,  we also included  the  fact that Cherenkov light has an opening angle of about $\alpha_{Cher.}=1.35^{\circ}$ for inclined directions, thus the Cherenkov light still can  hit both   mirror  and camera even if its direction  is outside  the 1.75$^{\circ}$  cone opened by the FOV of the MAGIC telescope. The following FOV  condition was used: $\alpha_{5 \mathrm{km}}= \arcsin(d_{5\mathrm{km}}/r_{5 \mathrm{km}}) < (\vartheta_{FOV}/2+\alpha_{\mathrm{Cher.}}) \approx 3.10^{\circ}$, where  $d_{5 \mathrm{km}}$ is the  distance  of the estimated shower  maximum to the shower axis and  $\vartheta_{FOV}=3.5^{\circ}$ is  the FOV of MAGIC camera.

\begin{figure*}[t!]
 \centering
 \includegraphics [width=0.49\textwidth,height=7cm]{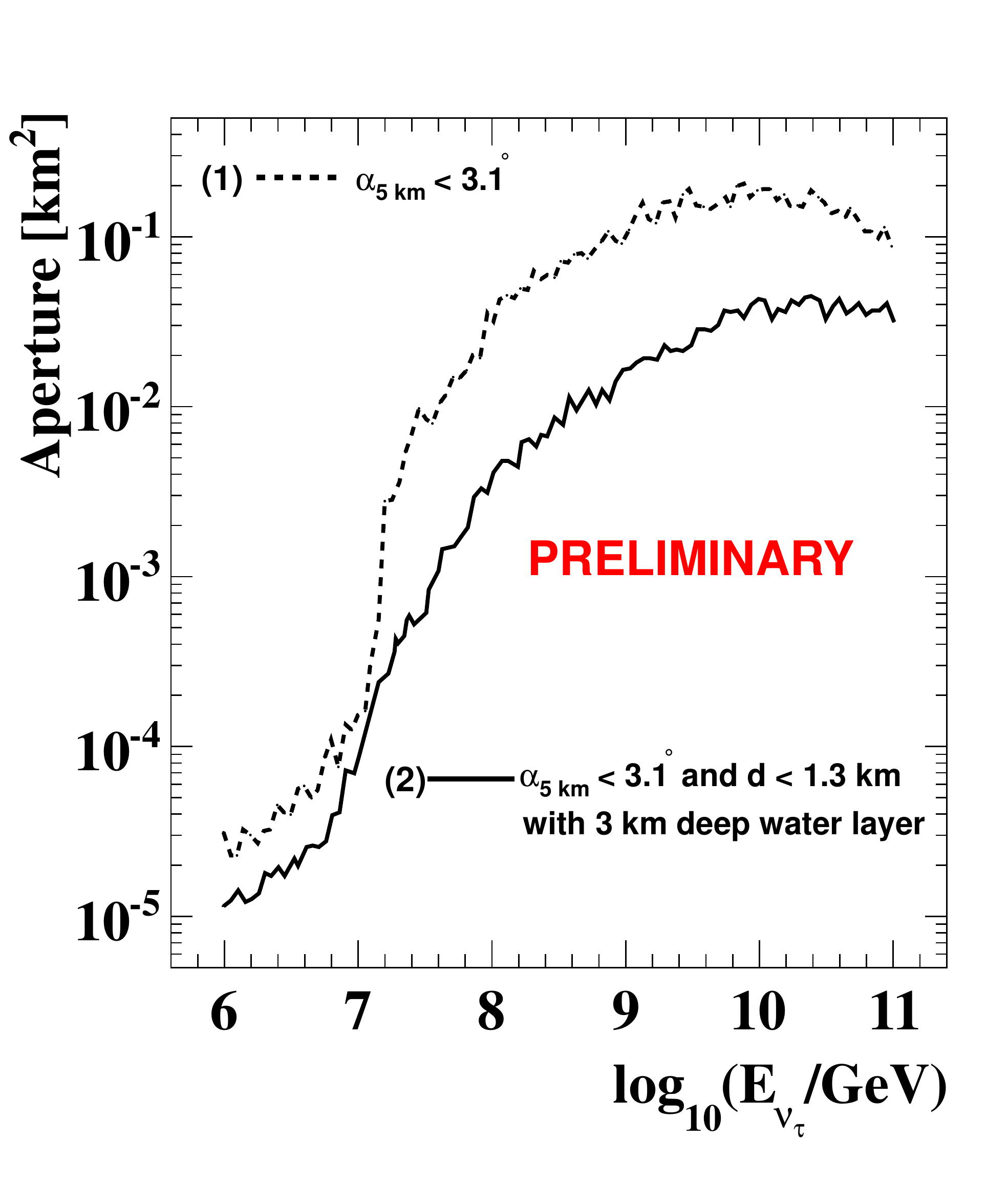}
  \includegraphics [width=0.48\textwidth,height=7cm]{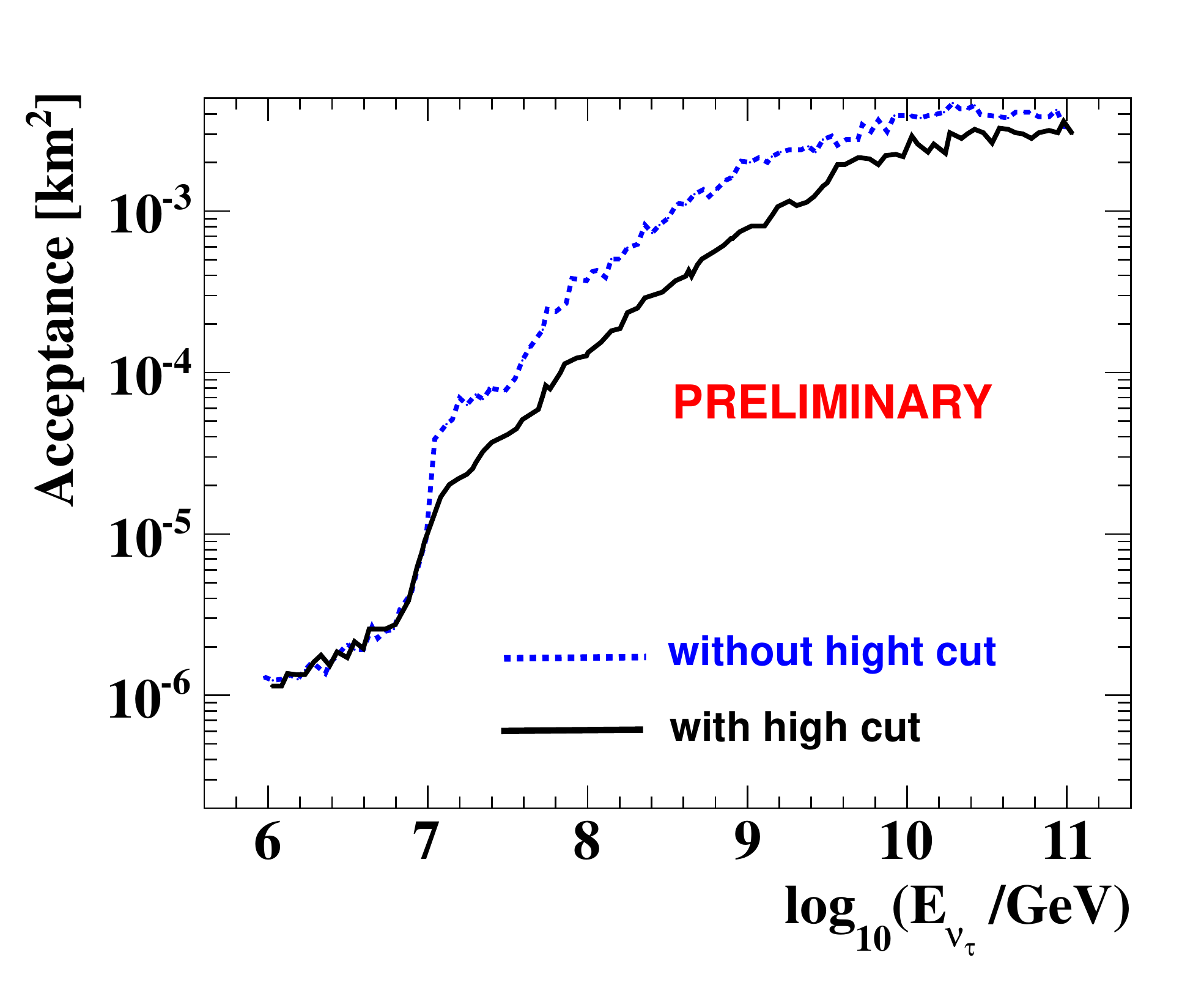}
 \caption{\small Left panel: The point-source aperture to  earth-skimming tau neutrinos for the MAGIC telescopes pointing at  $\theta_{\mathrm{MAGIC}}=92.5^{\circ}$ and  $\phi_{\mathrm{MAGIC}}=-30^{\circ}$. Computations including the orographic conditions of the MAGIC site and for different densities of the interaction medium are shown (see text for more details); Right panel: MC acceptance for point sources, $A^{\mathrm{PS}}(E_{\nu_\tau})$, for  earth-skimming tau neutrinos,  as estimated for the MAGIC site, with a FOV cut of: $\alpha_{5\mathrm{km}}<3.1^{\circ}$ and $d < 1.3$ km and within  identification efficiencies shown in \cite{tauicrc2017}. Note
 the significant influence of the heigh cut on the calculated acceptance.}\label{fig2}
\vspace{-0.25cm}
\end{figure*}

In Figure~\ref{fig2} (left panel)   we show  an estimate  of  the MAGIC point-source aperture (for $T_{\mathrm{eff},i}=1$) to tau neutrinos.
The aperture is shown for  two  cases: (1) for  simulations including  the orography of the La Palma island, 
but with the  spherical model of  Earth,  with the rock density of about 2.65 g/cm$^2$, outside the island, and 
 with the $\alpha_{r_{5 \mathrm{km}}}<3.1^{\circ}$  cut;  (2)  for  simulations including  the orography of the La Palma island \cite{dem} and the 3~km deep water layer around the La Palma island.  As we can see in the plot, the water layer is important because it leads to about  a factor two (in the energy range 10-100 PeV) smaller aperture than for the spherical Earth calculations with a rock density of about 2.65 g/cm$^2$ outside the La Palma island. This is because the mixture of a first dense material (rock) and then much less dense one produces many tau leptons which then decay inside the water and are lost for observations.  
Another important effect, which was also  included in our calculations are possible clouds during observations.
We  can have  information  from the MAGIC lidar  system~\cite{lidar}, 
whenever high clouds in the  vertical direction are present at the MAGIC site. However,  we did  not have any information about clouds present in the direction of the  seaON and seaOFF observations, due to the lack of possible data measurement
in these directions. Thus, in our acceptance calculations we  include the quasi-stable sea of cumulus between 1500 and 1900 m a.s.l., usually present  at the MAGIC site due to the temperature inversion layer at these altitudes. To estimate the effect of this we assumed that all decaying tau leptons  below 1500 a.s.l. are discarded in the  acceptance  calculations i.e. in $N_{\mathrm{FOVcut}}$. We assumed that  for such a case  the  Cherenkov light is absorbed  when it passes the layer from 1.5-1.9 km a.s.l. Such a cut provides  a  conservative  upper limit on this effect. In Figure~\ref{fig2} (right panel)   we  show the acceptance, when our selection cut, identification efficiency and the  height cut is  included in our simulations.
As we can see in the  plot  the height cut  leads  also to a smaller (about factor two) acceptance.  
 \section{Event rate and tau neutrino sensitivity}
\begin{figure*}[t!]
 \centering
 \includegraphics[width= 0.45\textwidth,height=6cm]{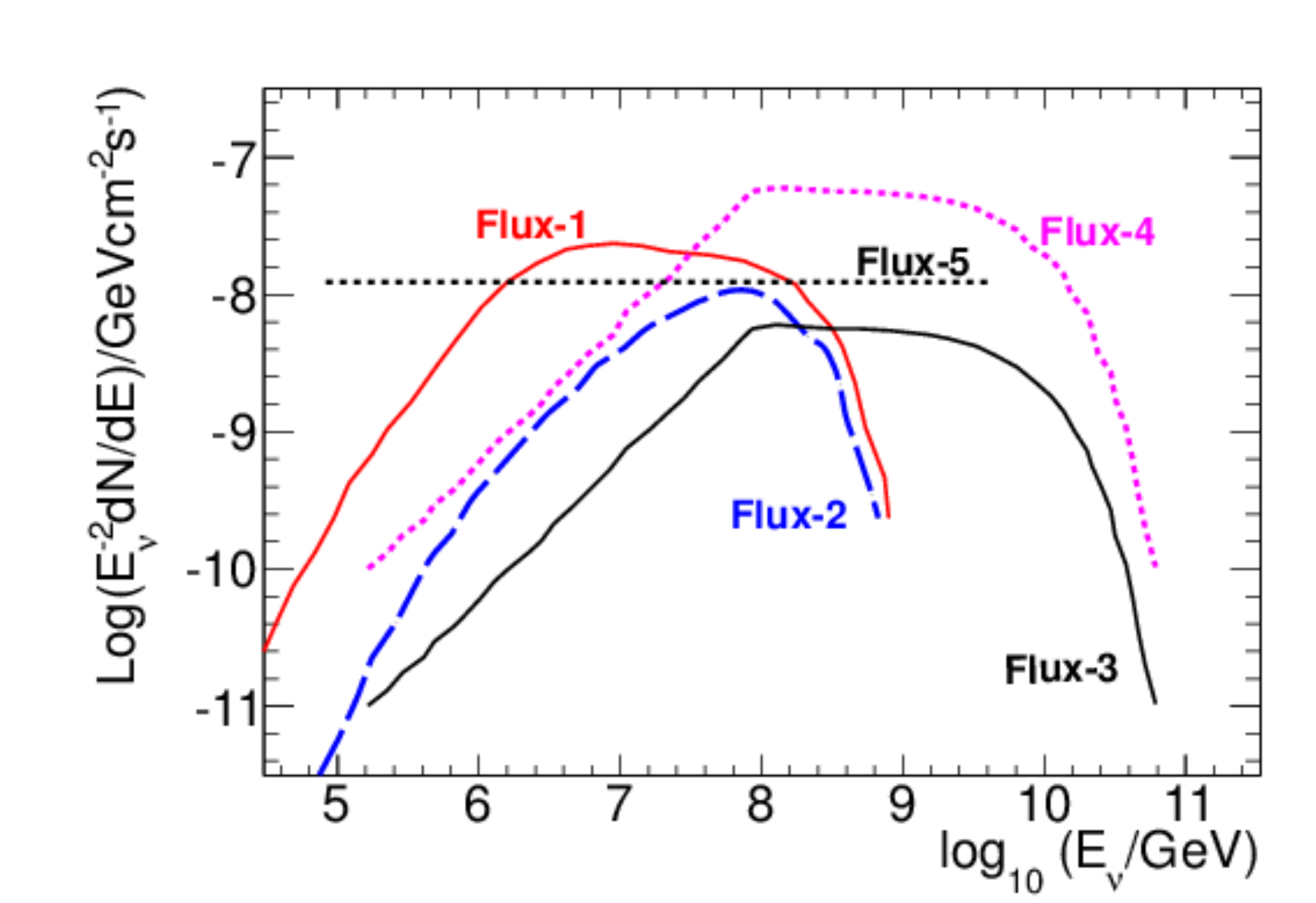}
 \includegraphics [width=0.45\textwidth]{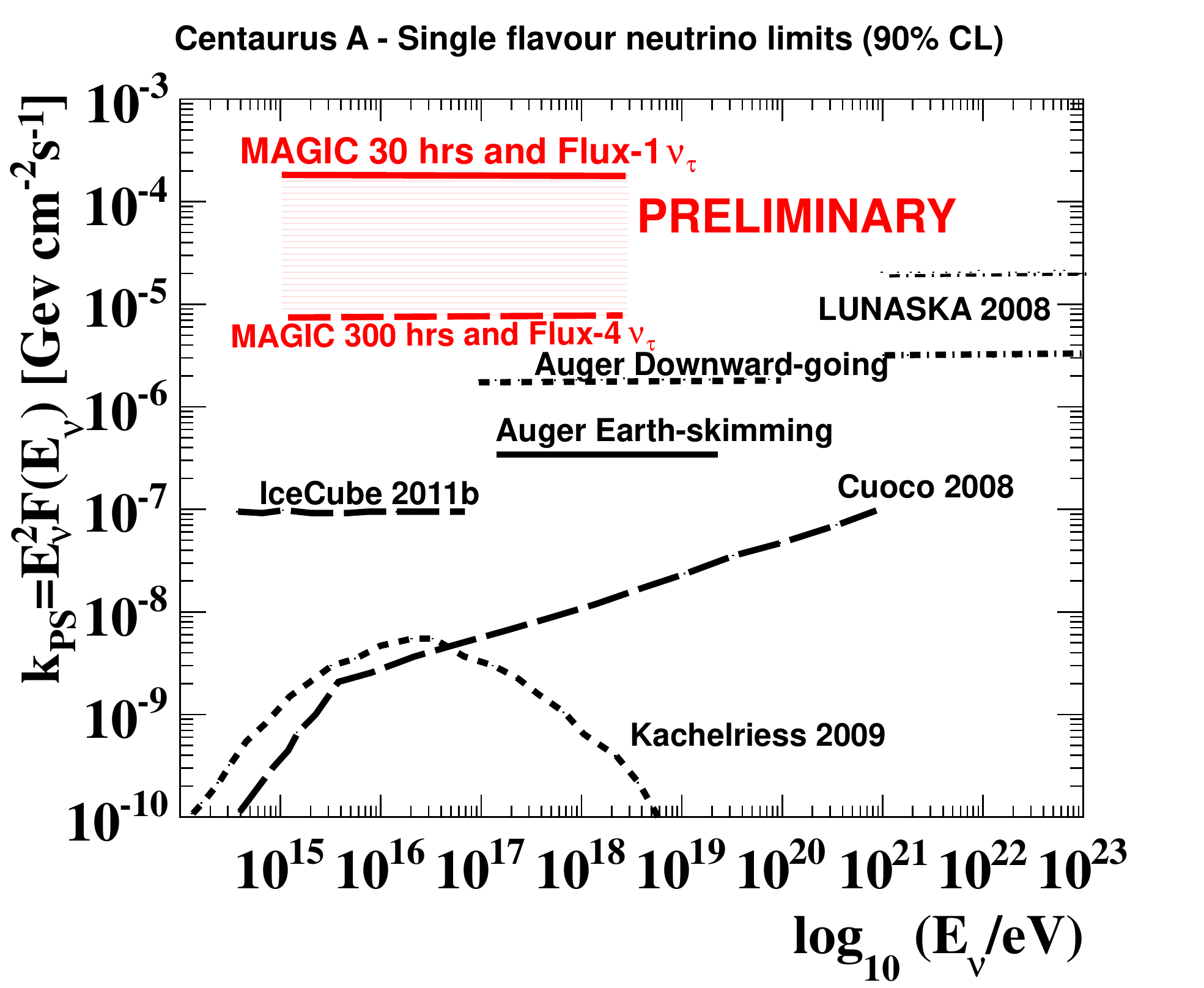}
 \caption{\small Left panel: A sample of representative neutrino flux predictions from photo-hadronic interactions in AGNs. Flux-1 and Flux-2 are calculations for the $\gamma$-ray flare of 3C 279~\cite{2009IJMPD}. Flux-3 and Flux-4 represent predictions for PKS~2155-304~\cite{Becker2011269}. Flux-5 corresponds to a prediction for 3C~279~\cite{PhysRevLett.87.221102}.
   Right panel:  The Pierre Auger upper limits at 90\% C.L. on a single flavor neutrino  flux from the active galaxy Centaurus A 
from the Earth-skimming and downward-going neutrino analysis \cite{auger}. 
 The MAGIC point-source sensitivity for 30/300 hours is marked by red solid  lines and the  hatched area. Note the different energy range for MAGIC and the Pierre
Auger Observatory.  Plot adopted from ~\cite{auger}.
  }\label{fig111}
\end{figure*}

Object like flaring sources, including GRBs, AGNs, Tidal disruption or the  Low Luminosity GRB (LLGRBS),
 can provide a boosted flux of neutrinos. Thus,  we provide an estimate of the event rate for a sample  of generic neutrino fluxes, from photo-hadronic interactions in case of flaring AGNs, if observed within the MAGIC observation window.
 In Table~\ref{table3} the expected event rates for MAGIC   are shown for fluxes from the  AGN benchmark models  shown  in Figure~\ref{fig111} (left panel). The rate is calculated for tau neutrinos   assuming that the source is in  the MAGIC  telescope FOV  for a period of 3 hours and with the acceptance calculated using the height cut/without height cut. For   Flux-3 and Flux-4   the event rate  is   at the  level of $3 \times 10^{-5}$ (i.e. those models covering the energy range beyond   $\sim 1\times10^{8}$ GeV  and after application of the height cut). For neutrino fluxes covering the energy range below $\sim 5\times10^{7}$\,GeV (Flux-1, Flux-2, Flux-5), the number of expected events  is  not larger than  $1.1 \times 10^{-5}$. In case of  no height cut  applied, the number of expected events is about a factor of two larger.
 
 \begin{table*}[b!]
\caption{\small Expected event rates for the MAGIC detector   in case of AGN flares. Case A: simulations with the
 height cut included, case  B without the height cut.}\label{table3}
\center
\small
\vspace{-0.2cm}
\begin{tabular}{cccccccccc}
\hline
\hline
& &\bf Flux-1  &\bf Flux-2& \bf  Flux-3 & \bf Flux-4 & \bf Flux-5 \\
& & \bf ($\times 10^{-5}/3$ hrs)   & \bf ($\times 10^{-5}/3$ hrs)  &  \bf ($\times 10^{-5}/3$ hrs)    &\bf ($\times 10^{-5}/3$ hrs)  &\bf ($\times 10^{-5}/3$ hrs)   \\
% &    &   &      &  & &\\
\hline
\hline
$N_{\mathrm{Events}}$& case A& 2.4 & 1.4 & 0.74   &7.4 &2.4\\
$N_{\mathrm{Events}}$&case  B  & 1.1&  0.6 &  0.30  & 2.9 & 1.2 \\
%$N^{\mathrm{Southern \mbox{ } Sky}}_{\mathrm{IceCube}}$& 1.18&  0.32 &  0.076  & 0.76 & 0.88 \\
\hline
\hline
\vspace{-0.7cm}
\end{tabular}
\end{table*} 
 
From the estimated acceptance with the height cut, the sensitivity for an injected spectrum $K\times\Phi(E_{\nu})$ with a known shape $\Phi(E_{\nu})$ was                                                                                                      
calculated. The 90\% C.L. on the value of $K$, according to ~\cite{limit} is $K_{90\%}=2.44/N_\mathrm{Events}$, with the assumption of                                                                                                                   
negligible background, zero neutrino events being observed by the MAGIC  during  sea observations, and  in case of an assumed flux of  $\Phi(E_{\nu})=1 \times 10^{-8} E^{-2} \mbox{ [GeV cm$^{-2}$ s$^{-1}$]} $,  the 90 \% C.L. limit for a point source search  is then:
\begin{equation}
 E_{\nu_\tau}^{2}\Phi^{PS}(E_{\nu_\tau}) < 2.0   \times 10^{-4} \mbox{ [GeV cm$^{-2}$ s$^{-1}$]} \label{limit}
\end{equation}
in the range from 2 to 1000 PeV. The  sensitivity is calculated  for  the  expected number of tau neutrino events equal to  $ N_\mathrm{Events}=1.2 \times10^{-4}$, based on the result listed in  Table~\ref{table3} for Flux-5, and   for  30 hours of observation time. The limit can be improved about one order of magnitude for larger observation times, such as the expected event rate in case of a strong flare. As it is seen from Table~\ref{table3}  for  Flux-4, and  the  observation time of $\sim 300$ hrs i.e.   the expected number  of tau neutrino events is $ N_\mathrm{Events}=2.9 \times10^{-3}$,  thus the 90\% C.L. limit could  reach  the value  of  $E_{\nu_\tau}^{2}\Phi^{PS}(E_{\nu_\tau}) < 8.4\times 10^{-6}$ \mbox{ [GeV cm$^{-2}$ s$^{-1}$]}  i.e. about  factor two lower than the limit of  the down-going analysis  of the Pierre Auger Observatory ~\cite{auger}, see Figure~\ref{fig111} (right panel). Note that the 
Auger downward limit corresponds  to an  equivalent  exposure  of  about $\sim 17520$ hrs, while MAGIC one 
(from Eq. \ref{limit}) to only 30 hrs. If we take into account  this difference then the MAGIC sensitivity for tau neutrinos  is at  the similar level like  for  the  the down-going point source analysis of the Pierre Auger Observatory i.e. if  observations will be  made for  the same number of hours.

 \section{Summary}
 We have presented results from MC simulations of $\tau$-induced air showers  and MAGIC observations at very high zenith angles. In particular, we have  calculated the point source  acceptance and the expected event rates,  for a sample  of generic neutrino fluxes from photo-hadronic interactions in AGNs.  Taking into account that for this purpose MAGIC has to be pointed below the horizon during moonless nights, the observational program for tau neutrino searches seems to be challenging, but in principle not impossible to pursue. In fact, a significant amount of  observation time can be accumulated  during periods with high clouds,  when those instruments are usually not used for gamma-ray observations. This means that during a few  IACTs observational seasons a significant amount of time  can be   accumulated. This makes the perspective for detection of tau neutrino induced showers by IACTs more attractive. With more than  1200 hours of data the achievable limit can reach the level of the results for down-going neutrinos  of the Pierre Auger Observatory. Finally, the  next-generation Cherenkov telescopes, i.e. the Cherenkov Telescope Array, will exploit their a larger FOV (in extended observation mode), and much larger effective areas.

\end{document}